\documentstyle[preprint,aps,epsf]{revtex}
\input epsf
\tighten
\draft
\begin{document}
\title{Rung-Rung Current Correlations on a 2-Leg t-J Ladder}

\author{Douglas J.~Scalapino\thanks{djs@vulcan.physics.ucsb.edu}}

\address{\sl Department of Physics,
         University of California,
         Santa Barbara, CA 93106-9530 USA}

\author{Steven R.~White\thanks{srwhite@uci.edu}}

\address{\sl Department of Physics,
         University of California,
         Irvine, CA 92697 USA}

\author{Ian~Affleck\thanks{affleck@physics.ubc.ca}}
\address{Canadian Institute for Advanced Research and Department of Physics
 and Astronomy, University of British Columbia, Vancouver,
BC,  Canada, V6T 1Z1}

\date{\today}
\maketitle

\begin{abstract}

We report the results of numerical calculations of rung-rung current
correlations on a 2-leg $t$-$J$ ladder with $J/t=0.35$ for dopings
$x=0.125$ and $x=0.19$. We find that the amplitude of these correlations
decays exponentially.  We argue that this can be understood within a
bosonization framework in terms of the pinned phase variables associated
with a C1SO phase with $d_{x^2-y^2}$-like power law pairing correlations.
\end{abstract}


\setlength{\baselineskip}{.3in}
\renewcommand{\thefootnote}{\fnsymbol{footnote}}

Variational flux-phase states are known to provide interesting low energy
estimates for the ground state of the 2D Heisenberg antiferromagnet.  In
addition, recent calculations have found signatures of a staggered-flux
phase in current vorticity correlations in lightly-doped $t$-$J$ clusters.
In particular, Ivanov {\it et.~al.} \cite{ILW00}
 found evidence for a power law decay of
the staggered current vorticity for a projected $d$-wave variational wave
function on a $10\times10$ lattice and Leung \cite{Leupp}
 has reported staggered
vorticity correlations from the exact diagonalization of a 32-site $t$-$J$
cluster with two holes.  Prompted by these results, we have carried out
density matrix renormalization group
(DMRG) calculations of the rung-rung current correlations on doped 2-leg
$t$-$J$ ladders.  As is known, near half-filling an isotropic $t$-$J$
ladder, with $J/t$ in the physical range appropriate to the cuprates,
retains a spin gap and exhibits power law $d_{x^2-y^2}$-like pair field and
$4k_F$-CDW correlations \cite{DR96}.  Thus, it is of interest to
explore whether there
may also be staggered current correlations on the 2-leg ladder.

Indeed, as discussed below, the DMRG calculations show that there are
oscillating rung-rung current correlations.  However, we find that they are
incommensurate at a finite doping, relatively
weak, and decay exponentially on a scale of several lattice spacings.
Complementing the numerical results, we discuss a bosonization treatment of
the 2-leg ladder. Here, one finds that the rung current depends exponentially
on various boson phase fields.
We argue that in a C1SO phase which has a spin-gapped ground state with
power law $d_{x^2-y^2}$-like pairing correlations, a phase variable
appearing in the rung-current is dual to a pinned variable and that this
leads to an exponential decay of the rung-rung current correlations, as seen
numerically.

Here we will consider an isotropic 2-leg $t$-$J$ ladder described by the
Hamiltonian
\begin{eqnarray}
H&=& t\sum_{i, \lambda, \alpha} \left(c^\dagger_{i+1\lambda\alpha}
c_{i\lambda\alpha} +
h.c.\right) - t\sum_{i,\alpha}
\left(c^\dagger_{i2\alpha}c_{i1\alpha}+h.c.\right)\nonumber\\
&-&J\sum_{i,\lambda}\left(\vec S_{i\lambda}\cdot \vec S_{i+1\lambda} -
\frac{n_{i\lambda}n_{i+1\lambda}}{4}\right)
- J\sum_i\left(\vec S_{i1}\cdot\vec S_{i2}-\frac{n_{i1}n_{i2}}{4}\right)
\label{one}
\end{eqnarray}
Here $c_{i\lambda\alpha}$ destroys an electron on rung $i$ and leg
$\lambda=1,2$ with spin $\alpha=\uparrow,\downarrow$ and $\vec
S_{i\lambda}=c^\dagger_{i\lambda}\ \vec\sigma/2\ c_{i\lambda}$. The Hilbert
space excludes all states with doubly-occupied sites. The rung current
operator for the $j^{\rm th}$ rung can be written as
\begin{equation}
J_j=it \sum_\alpha \left(c^\dagger_{j2\alpha}c_{j1\alpha} -
c^\dagger_{j1\alpha} c_{j2\alpha}\right) = it \sum_\alpha
\left(\psi^\dagger_{je\alpha} \psi_{jo\alpha}-\psi^\dagger_{jo\alpha}
\psi_{je\alpha}\right)
\label{two}
\end{equation}
Here the second form is expressed in terms of the even (bonding) and odd
(antibonding) operators
\begin{equation}
\psi_{je/o\alpha}= \frac{c_{j2\alpha} \pm c_{j1\alpha}}{\sqrt{2}}
\label{three}
\end{equation}
with $e$ and $o$ corresponding to $+$ and $-$ respectively. We will study
the ground state rung-rung correlation function
\begin{equation}
c(|k-j|) = \left\langle J_k J_j\right\rangle
\label{four}
\end{equation}

Using DMRG techniques \cite{Whi93} we have calculated the rung-rung
current correlation
function, Eq.~(\ref{four}), for a $2\times 32$ ladder with $J/t=0.35$ at
fillings $\langle n_i\rangle = 1-x$ with $x=0.125$ and $0.19$. The DMRG
results are for a ladder with open-end boundary conditions and $c(\ell)$ is
obtained by averaging-over sites separated by $\ell=|k-j|$.
The calculations were done with the finite system DMRG method,
keeping a maximum of 800 states per block, on a $32\times2$ open
system. The discarded weight was typically about $3\times10^{-6}$ in the
final sweep. The correlation measurements were checked by repeating
some of the calculations keeping 1000 states;
the differences were quite small and would not affect any
conclusions.
In order to reduce the effect of the open boundaries,
results for many
different points with the same separation were averaged over to
obtain each data point shown.
As shown in Fig.~1(a) for $x=0.125$, the rung-rung current
correlations exhibit an incommensurate oscillation with a small amplitude
and decay exponentially. Fig.~1(b) shows a semi-log plot of
$|c(\ell)|$ versus $\ell$ for $J/t=0.35$ and $x=0.125$ and Fig.~1(c) shows a
similar plot for $x=0.19$. The slope of the solid line in Fig.~1(b) is $1/2.8$
corresponding to an exponential decay of $c(\ell)$ with a correlation
length $\xi=2.8$ lattice spacings.  Similarly, as seen in Fig.~1(c),
for $x=0.19$ we find an exponential decay of the rung-rung current
correlations with $\xi=4$ lattice spacings.

Further insight into the nature of rung-rung current correlations on a
2-leg $t$-$J$ ladder is provided within a bosonization framework.  Here we pass to
a continuum limit along the length of the 2-leg ladder and introduce the
usual left- and right-moving fields
\begin{equation}
\psi_{\lambda\alpha}(x)=e^{-ik_{F\lambda}x} \psi_{L\lambda\alpha}(x) +
e^{ik_{F\lambda}x} \psi_{R\lambda\alpha}(x)
\label{five}
\end{equation}
with $\lambda=e$ and $o$ corresponding to the even (bonding) and odd
(antibonding) bands respectively with $k_{Fe}$ and $k_{Fo}$ the fermi wave
vectors for these bands.

Using this representation, the rung current operator, Eq.~(\ref{two}),
becomes
\begin{equation}
J(x) = J_o (x) + \left(e^{i\left(k_{Fe}+k_{Fo}\right)x}\, J_{2k_F} (x) +
h.c.\right)
\label{six}
\end{equation}
with
\begin{equation}
J_o (x) = it \sum_\alpha \biggl[ e^{i\left(k_{Fe}-k_{Fo}\right)x}
\, \left(\psi^\dagger_{Le\alpha}(x) \psi_{Lo\alpha} (x) -
\psi^\dagger_{Ro\alpha} (x) \psi_{Re\alpha} (x)\right) +
h.c.\biggr] \label{seven}
\end{equation}
and
\begin{equation}
J_{2k_F}á= it \sum_\alpha \left(\psi^\dagger_{Le\alpha}(x)
\psi_{Ro\alpha}(x) - \psi^\dagger_{Lo\alpha}(x) \psi_{Re\alpha}(x)\right)
\ .
\label{eight}
\end{equation}
In the usual way, we represent the left and right moving fermion fields
by left and
right moving boson fields:
\begin{equation}
\psi_{L/R\lambda \alpha}\sim e^{i\sqrt{4\pi}\phi_{L/R\lambda
\alpha}}
\end{equation}
Then, introducing the dual canonical Bose fields,
\begin{equation}
\phi_{\lambda\alpha}= \phi_{R\lambda\alpha} + \phi_{L\lambda\alpha}
\qquad \theta_{\lambda\alpha} = \phi_{R\lambda\alpha}-\phi_{L\lambda\alpha}
\label{nine}
\end{equation}
we have, for example, for the first term in Eq.~(\ref{eight})
\begin{equation}
\psi^\dagger_{Le\alpha}(x) \psi_{Ro\alpha} (x) \sim e^{i\sqrt{\pi}
\left(-\phi_{e\alpha}+\theta_{e\alpha} + \phi_{o\alpha} +
\theta_{o\alpha}\right)}\ .
\label{ten}
\end{equation}
In terms of the even and odd parity charge and spin Bose fields with
$\lambda=e$ or $o$
\begin{eqnarray}
\phi_{\lambda\rho}&=& (\phi_{\lambda\uparrow} +
\phi_{\lambda\downarrow})/\sqrt{2}\nonumber\\
\phi_{\lambda\sigma} &=& (\phi_{\lambda\uparrow} - \phi_{\lambda\downarrow}
)/\sqrt{2}
\label{eleven}
\end{eqnarray}
and their dual fields, we can write the up spin part of Eq.~(\ref{ten}) as
\begin{equation}
\psi^\dagger_{Le\uparrow} (x) \psi_{Ro\uparrow} (x) \sim e^{i\sqrt{\pi}
\left(-\phi_{-\rho} - \phi_{-\sigma} - \theta_{+\rho} -
\theta_{+\sigma}\right)}\ .
\label{twelve}
\end{equation}
Here,
\begin{equation}
\phi_{\pm\rho} = \left(\phi_{e\rho} \pm \phi_{o\rho}\right)/\sqrt{2}
\label{thirteen}
\end{equation}
with similar relations for $\phi_{\pm\sigma}$, $\theta_{\pm\rho}$, and
$\theta_{\pm\sigma}$. Actually, this last transformation is not
canonical when the even and odd bosons have different velocities.
However, we will follow the standard practice \cite{Fab93,BF96}
 of assuming that this
velocity difference is irrelevant.
In this way, the rung current operators, Eqs.~(\ref{seven}) and
(\ref{eight}) become
\begin{equation}
J_o(x)\sim i\sum_{\delta_1,\delta_2= \pm 1}\delta_1e^{i\pi
(k_{Fe}-k_{Fo})x}
e^{-i\sqrt{\pi}[\phi_{-\rho}+\delta_2\phi_{-\sigma}-
\delta_1\theta_{-\rho} -\delta_1\delta_2\theta_{-\sigma}]}+h.c.
\label{fourteen}\end{equation} (where $\delta_1$ labels the left
or right term and $\delta_2$ labels spin) and
\begin{equation}
J_{2k_F} (x) \sim \sum_{\delta_1,\delta_2=\pm 1}\delta_1
e^{i\sqrt{\pi}[-\delta_1\phi_{-\rho}
-\delta_1\delta_2\phi_{-\sigma}+\theta_{+\rho}+\delta_2\theta_{+\sigma}}]
\label{fifteen}
\end{equation}
(where now $\delta_1$ labels the $e-o$ or $o-e$ terms and $\delta_2$
labels spin).

Renormalization group analysis \cite{Fab93,BF96}, based on the weak
coupling 2-leg Hubbard
model and earlier numerical results \cite{Dagotto,Poil,Whi97}
on $t$-$J$ ladders suggest that, for a
realistic parameter range away from half-filling, the $t$-$J$ model is in a
``C1SO'' phase in which both spin bosons are gapped and one of the charge
bosons is gapped.  When a boson is gapped, either $\langle \phi \rangle
\not= 0$ or $\langle \theta \rangle \not= 0$. These expectation values must
be specified to completely specify the phase which the system is in.  The
expected phase has
\begin{equation}
\left\langle\theta_{\pm\sigma} \right\rangle \not= 0\ , \quad
\left\langle\phi_{-\rho}\right\rangle\not=0\ .
\label{sixteen}
\end{equation}
In this phase the uniform part of the pair correlation function has
power-law decay and the $2k_{Fe}+2k_{Fo}$ part of the density correlation
function has power law decay (but not the $2k_{Fe}$ or $2k_{Fo}$ parts),
in agreement with the numerical results.

We may replace factors like $e^{i\sqrt{\pi}\phi_{-\rho}}$ by their
expectation values, so that:
\begin{equation}
J_{2k_F} \sim \sum_{\delta_1, \delta_2 =\pm}\delta_1
e^{i\sqrt{\pi}\left[-\delta_1\delta_2\phi_{-\sigma} +
\theta_{+\rho}\right]}\ . \label{seventeen}
\end{equation}
Correlation functions involving exponentials of $\phi_{-\sigma}$ decay
exponentially since $\phi_{-\sigma}$ is conjugate to the pinned dual field
$\langle\theta_{-\sigma}\rangle$.  Therefore, we expect that the term in
the rung current correlation function which oscillates at $2k_F$ will decay
exponentially.  Similarly, since $J_o$ depends exponentially on
$\phi_{-\sigma}$ and $\theta_{-\rho}$ and both of these are conjugate to
pinned phases, these correlation function will also decay
exponentially.

Based upon this bosonization analysis, we expect that the
asymptotic form of the rung-rung current correlations will
oscillate with an incommensurate wave vector $(k_{Fo} + k_{Fe}) =
\pi (1-x)$. Thus, for $x=0.125$ the bosonization result gives
\begin{equation}
c(\ell) \sim e^{-\ell/\xi} \cos (7\pi\ell/8+\phi)
\label{bosres}
\end{equation}
Using the DMRG results, $\xi=2.8$ for $x=0.125$, Fig.~2 shows a comparison
of Eq.~(\ref{bosres}) with the DMRG data.

We note that other phases have been suggested for the 2-leg Hubbard model
and extended Hubbard models
\cite{Fab93,Vojta}
some of which could have power law decay for the $2k_F$ component of the rung
current correlation function.  
However, power law decay could only occur in a phase in which
none of the boson fields dual to the ones appearing in Eq. (16)
are pinned.  This would mean that each boson field in Eq. (16)
is either itself pinned or else is gapless.  A possible C1S0
phase with this property has:
\begin{equation}
\left\langle\theta_{+\sigma}\right\rangle \not= 0\ , \quad
\left\langle\phi_{-\sigma}\right\rangle \not=0\ ,\quad
\left\langle\phi_{-\rho}\right\rangle\not=0\ .
\label{eighteen}
\end{equation}
In fact, such a C1S0  phase was predicted by Fabrizio \cite{Fab93}
in the Hubbard model for small
$t_\perp$ and a small range of $U$ near 7, based on a perturbative
renormalization group analysis.
Thus, it is possible that such a phase might also
occur in the 2-leg $t$-$J$ model or some generalization of it,
 for some range of parameters.  However,
Eq.~(\ref{eighteen}) is not a sufficient condition for power law decay.
Not only must these fields have non-zero ground state expectation values,
but also these values must be such that a cancellation of the power-law
part of the correlation function between the various terms in
Eq.~(\ref{fifteen}) does not occur.  By explicitly writing pairing operators
in the $\phi /\theta$, $\rho /\sigma$, $+/-$ basis, used above,
it can be
checked that {\it all} pair correlations (singlet and triplet) have
exponential decay in such a putative phase.
We emphasize again that, based upon the
$d_{x^2-y^2}$-like power law pairing correlations and density power law
correlations which are observed for the
$t$-$J$ ladder for physically relevant $J/t$ and doping regimes of
interest, it would
appear that the ground state of this model is
characterized by pinned phases given by
Eq.~(\ref{sixteen}).
In this ground state the current-current correlations decay
exponentially.

\acknowledgments

We acknowledge useful discussions with P.A.~Lee and J.B.~Marston regarding the
occurance of orbital antiferromagnetic currents.  DJS and IA
acknowledge the hospitality of the ITP, University of California, Santa
Barbara where this work was initiated.
DJS acknowledges support from the NSF under
grant No.~PHY99-07949 (ITP) and
grant No.~DMR98-17242 (DJS),  SRW acknowledges support from the NSF
under grant No.~DMR98-70930 and IA acknowledges support
from NSERC of Canada.

\begin{figure}
\epsfxsize=2.8 in\centerline{\epsffile{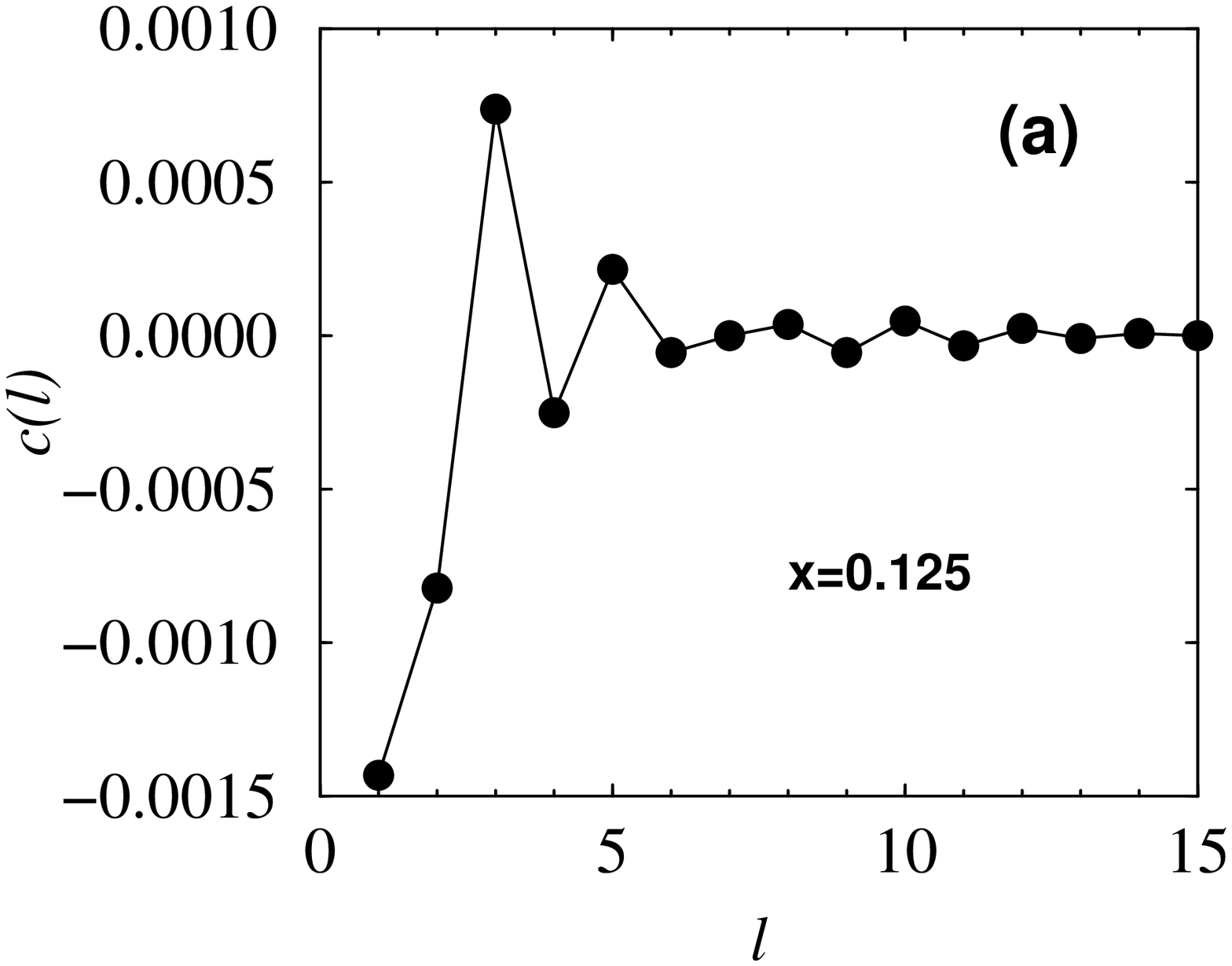}}
\epsfxsize=2.8 in\centerline{\epsffile{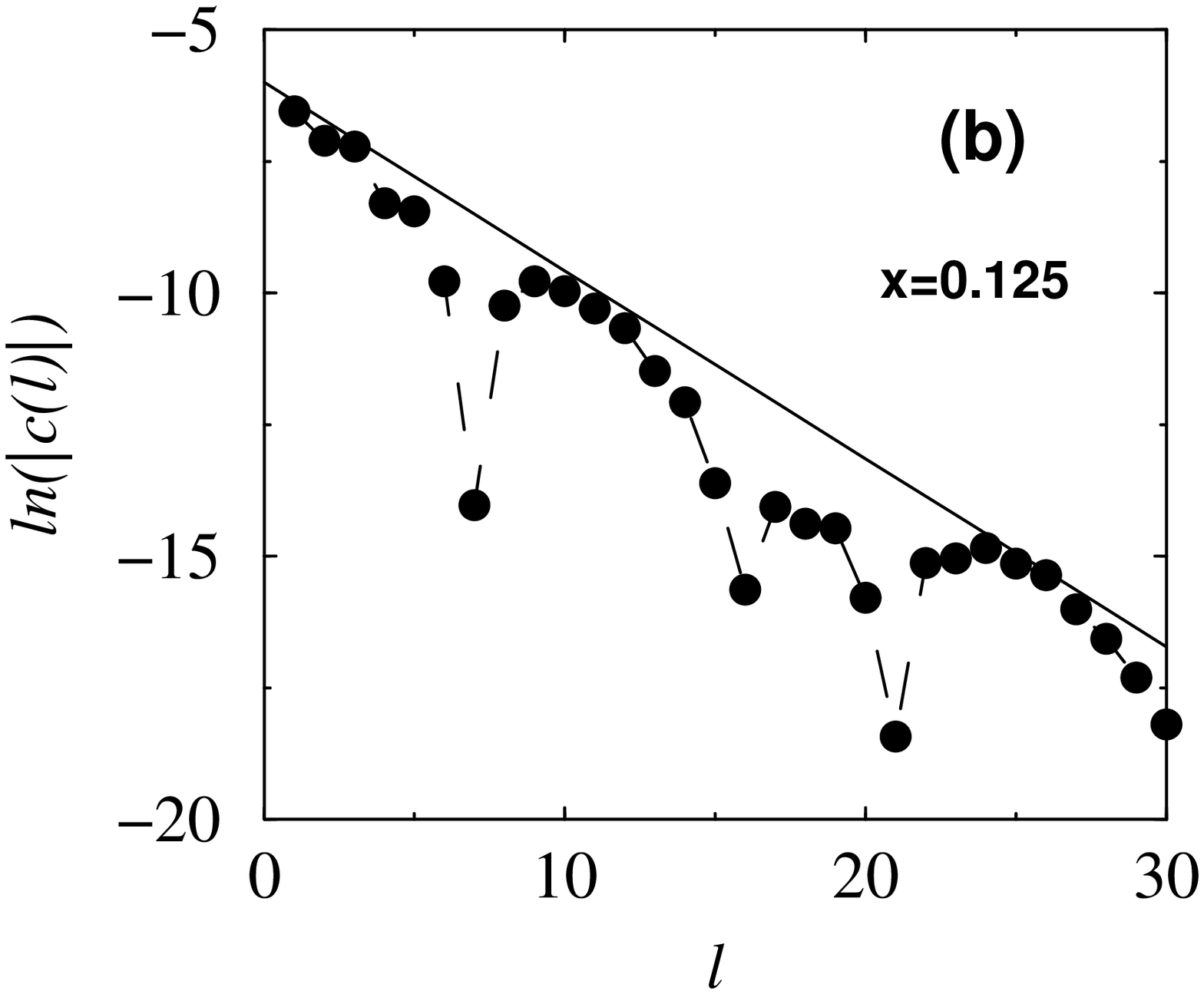}}
\epsfxsize=2.8 in\centerline{\epsffile{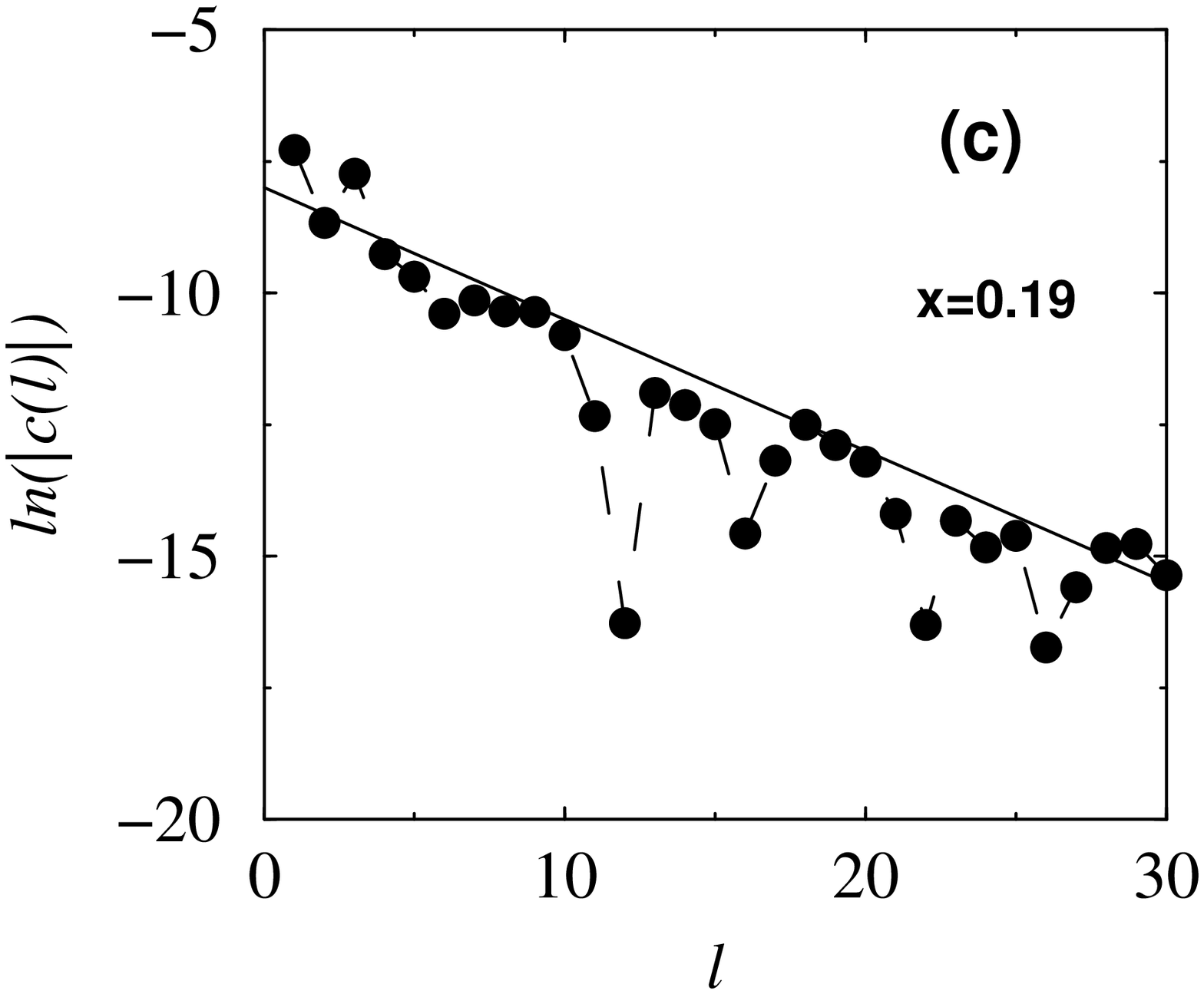}}
\caption{The current-current correlation function $c(l)$ on a $32\times2$ lattice
calculated with DMRG, at dopings of $x=0.125$ and $x=0.19$, with
$J/t=0.35$.
In (a), we show results on a linear scale for $x=0.125$, and
in (b), we show the same results on a semi-log scale. The solid line in (b)
corresponds to a decay length of $\xi=2.8$. In (c), we show
semi-log results for $x=0.19$. The solid line corresponds to a
decay length of $\xi=4$.}
\end{figure}
\begin{figure}[ht]
\epsfxsize=3.5 in\centerline{\epsffile{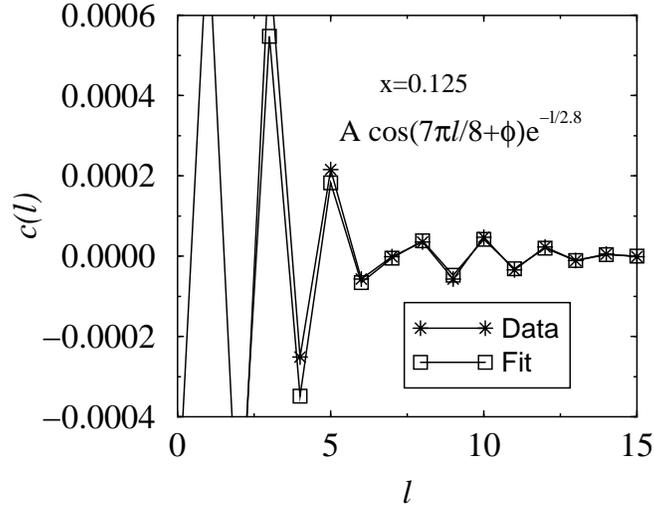}}
\caption{(a) The current-current correlation function $c(l)$ data from
Fig. 1(a) is compared with the asymptotic form expected from
bosonization. Here $A=0.0016$, $\phi = -2.0$, and as in Fig. 1(b)
the decay length is $\xi=2.8$ lattice spacings.}
\end{figure}

\end{document}